\documentclass[a4paper,11pt]{article}

\usepackage[utf8]{inputenc}
\usepackage{geometry}
\usepackage{fancyhdr}
\usepackage{setspace}
\usepackage{hyperref}
\usepackage{graphicx}
\usepackage{endnotes}
\let\footnote=\endnote

\usepackage[style=apa,backend=biber]{biblatex}
\begin{document}

\begin{center}
    {\Large{How permanent are metadata for research data? Understanding changes in DataCite metadata}}
\end{center}

\vspace{0.5cm}

\begin{itemize}
\centering
    \item[] Dorothea Strecker\textsuperscript{1} (corresponding author)\\dorothea.strecker@hu-berlin.de\\
    \href{https://orcid.org/0000-0002-9754-3807}{0000-0002-9754-3807}
\end{itemize}

\begin{center}
    $^1$ Humboldt-Universität zu Berlin, Berlin School of Library and Information Science
    \\
    \hspace{1cm}
    \\
    December 13, 2025
\end{center}

\vspace{1cm}

\section*{Abstract}
With the move towards open research information, the DOI registration agency DataCite is increasingly used as a source for metadata describing research data, for example to perform scientometric analyses.
However, there is a lack of research on how DataCite metadata describing research data are created and maintained.
This paper adresses this gap by using DataCite metadata provenance information to analyze the overall prevalence and patterns of change to DataCite metadata records.
Metadata change was observed for 12.18 \% of metadata records in the sample, and change tends to be incremental and not extensive. DataCite metadata records offer reliable descriptions of datasets and are stable enough to be used in scientometric research.
The rate of change differs from previous studies of metadata change in other contexts, suggesting that there are differences in metadata practices between research data repositories and more traditional cataloging environments.
The observed changes do not seem to fully align with idealized conceptualizations of metadata creation and maintenance for research data. In particular, the data does not show that metadata records are maintained routinely and continuously. Metadata change also has a limited effect on metadata completeness.

\section*{Keywords}
research data ; metadata ; metadata change ; open research information

\newpage
\onehalfspacing
\pagenumbering{arabic}

\section{Introduction}
DOI registration agencies like Crossref are important actors in the scholarly information ecosystem \parencite{hendricks_crossref_2020}. By allocating DOI prefixes and registering DOIs, they enable the persistent identification and reference of information resources.\footnote{doi Foundation: \url{https://www.doi.org/the-community/what-are-registration-agencies/} (accessed: 2024-12-03)}
With the move towards open research information, as exemplified by initiatives such as the Barcelona Declaration on Open Research Information\footnote{Barcelona Declaration on Open Research Information: \url{https://barcelona-declaration.org/commitments/}}, their function as metadata providers is becoming more prominent: DOI registration agencies aggregate and enrich metadata from distributed sources, and their metadata collections are used in research and form the basis of other services \parencite{van_eck_crossref_2024}.
DataCite was established in 2009 as a DOI registration agency specifically for research data \parencite{brase_approach_2009}.
Since then, the service has evolved beyond purely registering DOIs for research data and is emerging as an important actor in the wider research data ecosystem \parencite{buys_datacites_2023}.
DataCite is a supporter of the Barcelona Declaration on Open Research Information and an important source for open research information.
\\
Despite the potential of DataCite metadata, recent analyses have revealed challenges associated with reusing it, particularly with regards to the lack of metadata completeness \parencite{robinson-garcia_datacite_2017,ninkov_datasets_2021,strecker_quantitative_2021}. This is not surprising, given that DataCite metadata are created in distributed environments, outside traditional cataloging contexts and often without the involvement of trained professionals. Under these circumstances, metadata practices can vary \parencite{hillmann_metadata_2008}. This was also observed for another DOI provider, Crossref \parencite{van_eck_crossref_2024}.
\\
In the context of research data, metadata creation is often conceptualized as a continuous task. This means that metadata records are expected to change over time, as descriptions of data are updated and improved \parencite{edwards_science_2011}. In this scenario, metadata records describing research data are changed in order to successively improve quality.
Understanding how DOI metadata records are maintained is the first step to address problems with metadata. DOI metadata are metadata records that are submitted to DOI registration agencies when a DOI is registered \parencite{doi_foundation_doi_2023}. These metadata records adhere to the metadata schema defined by the DOI registration agency. DOI metadata records for research data are created and changed in a distributed environment by repositories with varying scopes and missions. Identifying patterns in metadata changes can provide insights into these complex workflows, and uncover approaches to improve metadata.
\\
However, there is a lack of research on metadata for research data \parencite{greenberg_big_2017,park_metadata_2010}. Investigations of processes related to the creation and revision of metadata for research data remain limited to small studies with a narrow scope \parencite{plantin_data_2018}. This research gap limits understanding of factors that result in incomplete metadata records for research data, and the identification of potential interventions to improve metadata quality.
\\
As will be discussed below, DOI metadata is often only one of several forms of metadata created at research data repositories, but its importance in the research data ecosystem is growing quickly. This paper therefore focuses on DataCite metadata. It traces how DataCite metadata records are created and maintained by analyzing metadata changes recorded by the service. Analyzing these change processes will contribute to a better understanding of DataCite metadata. To support the shift towards open research information, open sources of research information need to be evaluated. A current example investigates metadata completeness in several sources of bibliographic data \parencite{delgado-quiros_completeness_2024}. Similar studies can contribute to a better understanding of DataCite as a source of open research information and advocate for its use and maintenance.
The permanence of metadata records is also of interest in other contexts where descriptions of research data are used. As an example, aggregators of metadata records might benefit from this information when planning their harvesting procedures. Potential reusers of datasets might also want to know how stable metadata records are over time -- if metadata records are generally mutable, they might question whether descriptions are permanent enough to base decisions regarding data reuse on.

\subsection{Research Questions}
This paper addresses the following research questions:
\begin{itemize}
    \item [RQ1] How common is change in DataCite metadata records describing research data?
    \item [RQ2] How do DataCite metadata records describing research data change over time?
\end{itemize}
In analyzing patterns of change in DataCite metadata records, this paper will show on a large scale how metadata describing research data are created and maintained.

\section{Background}
The following section will provide a general overview of research on metadata for research data and practices at research data repositories, before summarizing current literature on metadata versioning and change.

\subsection{Metadata for research data}
\begin{figure}
    \centering
    \includegraphics[width=0.9\linewidth]{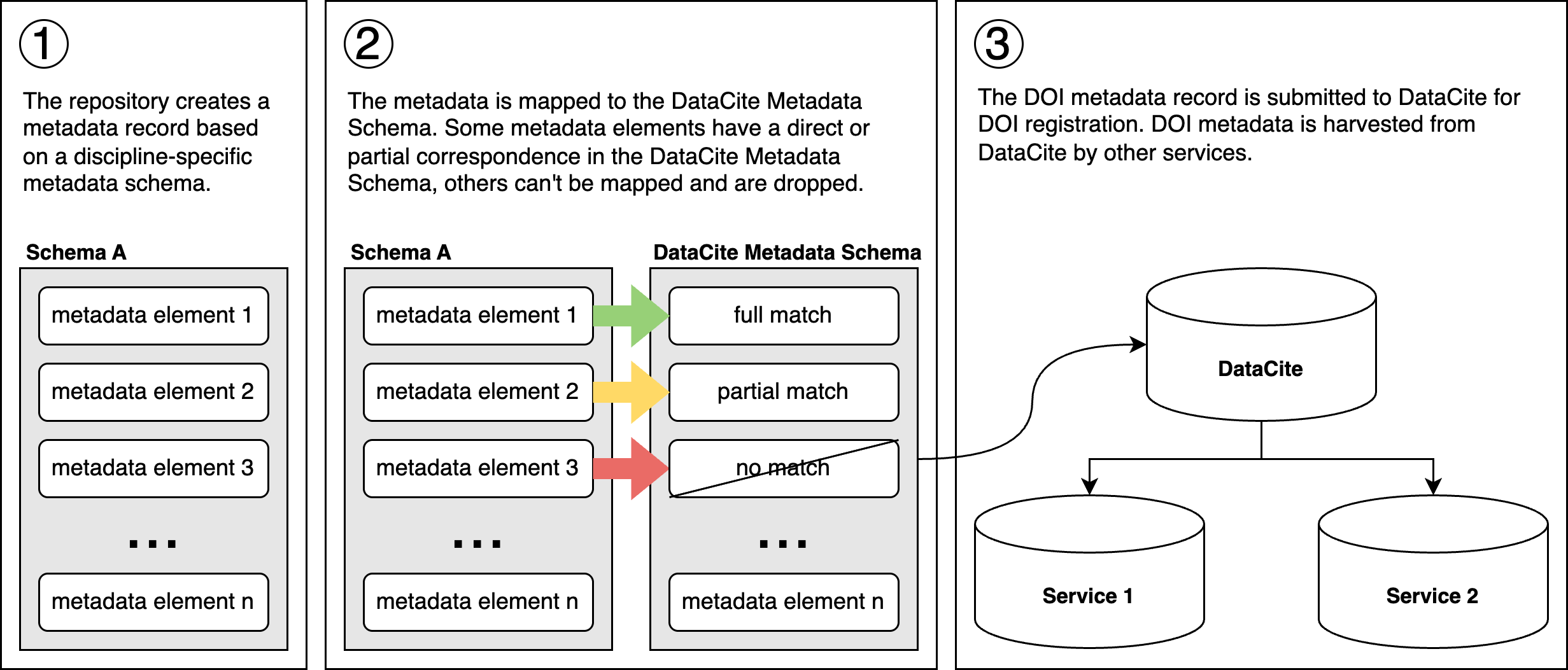}
    \caption{Schematic overview of an example metadata workflow.}
    \label{fig:metadata_workflow}
\end{figure}

Metadata are "structured, encoded data that describe the characteristics of information-bearing entities." \parencite[12]{zeng_metadata_2022} They take the form of statements about the information-bearing object they describe \parencite{pomerantz_metadata_2015}. The types of statements that can be made are defined by a metadata schema \parencite{zeng_metadata_2022}. All statements about an information-bearing object are referred to as a \emph{metadata record} that comprises a set of \emph{metadata elements} and their \emph{values} \parencite{enoksson_activity_2015}.
Metadata enable core functionalities of information infrastructures \parencite{baca_setting_2008,riley_understanding_2017}, such as discovery, access, and preservation.
\\
In the context of research data, creating structured metadata is even more important than for textual resources, because data are not self-explanatory, and in most cases, there is no text available to extract semantic information from \parencite{hansson_open_2022}. There are numerous metadata schemas available for describing research data \parencite{willis_analysis_2012}, and research data repositories vary in the schemas they choose to describe their collections \parencite{asok_common_2024,koshoffer_giving_2018}. Even within a relatively narrow field, the use of metadata standards is not uniform \parencite{mayernik_role_2022}.
\\
Research data repositories face tensions because use scenarios require both general and homogenous metadata, for example to facilitate the development of comprehensive discovery services spanning multiple disciplines, and domain specific metadata, for example to allow users to make sense of data and come to a decision about data reuse \parencite{de_vries_flexible_2022,doran_reconciling_2021,radio_manifestations_2017,sostek_discovering_2024}.
To resolve this tension, many research data repositories use multiple metadata schemas simultaneously for different purposes, for example a general schema such as the DataCite Metadata Schema for DOI registration, and a more specialized schema for local users' needs \parencite{habermann_connecting_2023,lee_landscape_2019,nose_enhancing_2024,wu_analysis_2023}.
\\
The DataCite Metadata Schema serves an integrating function in the wider research data ecosystem, because it is widely used by research data repositories, and it is general enough to be applicable to any discipline \parencite{mayernik_role_2022}. As a result, it brings together diverse data collections at one central service \parencite{hayslett_data_2015}.
Not all researchers register DOIs when publishing datasets \parencite{jiao_data_2022}. However, persistent identifiers and the metadata attached to them are essential to facilitate data publication and reuse \parencite{lee_developing_2014}. When a DOI is registered with DataCite, metadata describing the information resource based on the DataCite Metadata Schema has to be submitted to the service (the schema includes elements like \emph{creator}, \emph{title} or \emph{publicationYear}).
DataCite also connects members’ collections with other types of research outputs and provides metadata for other services. For example, DataCite is one of the largest sources for the general-purpose data discovery service Google Dataset Search: DataCite provides metadata for a large number of repositories that don’t expose their metadata in the formats required for indexing \parencite{benjelloun_google_2020}.
DataCite currently holds the most comprehensive collection of metadata for research data, and therefore is emerging as a data source for scientometric research \parencite{robinson-garcia_datacite_2017,sixto-costoya_exploring_2021}. The service will likely gain in importance in this area as the Data Citation Corpus, a collection of citation information for research data developed by DataCite, the Wellcome Trust and the Chan Zuckerberg Initiative, continues to evolve \parencite{vierkant_wellcome_2023,puebla_building_2024}.
\\
At research data repositories, metadata records are often created first based on a locally implemented, more specific metadata schema; they are then mapped to a more general schema for harvesting, DOI registration and other, more global purposes \parencite{habermann_connecting_2023,nose_enhancing_2024,taylor_think_2022}. However, in these transformation processes, semantic information from the original description can be lost, for example because of incongruent structures of metadata schemas \parencite{radio_manifestations_2017,taylor_think_2022}. Therefore, information that is available locally on the repository landing page might not be submitted to DataCite, resulting in diminishing completeness as metadata records are transferred to increasingly global settings \parencite{mayernik_role_2022}. Figure \ref{fig:metadata_workflow} depicts an example metadata workflow.

\subsection{Metadata change}
Research data repositories must find a balance between fixity and fluidity of their collections: On one hand, repositories must ensure the long-term availability, reliability and authenticity of datasets; on the other hand, research data are mutable objects and often, some degree of change is inevitable \parencite{daniels_managing_2012}.
\\
The same holds true for metadata: Metadata records create continuity in an information infrastructure, but they are also subject to change.
Creating metadata for research data is conceptualized as a continuous process that can never be considered truly complete or final, because metadata records are created from a specific perspective for a specific use \parencite{mayernik_tracing_2018}. If the broader environment of a repository or its user base shifts over time, metadata records must be adapted to meet new requirements \parencite{aspray_talking_2019,donaldson_data_2020,habermann_metadata_2018,habermann_connecting_2023,mayernik_modernizing_2017}. Metadata records can be revised and improved at several stages of repository workflows, even after a dataset is published \parencite{greenberg_metadata_2013,habermann_connecting_2023,thomer_craft_2022}. Metadata change can occur to reduce friction potential data reusers might face, for example after long time periods have passed since a dataset was first created \parencite{edwards_vast_2013}. Another important reason for metadata change is to add persistent identifiers to existing metadata records as the landscape of persistent identifiers is evolving \parencite{habermann_connecting_2023}. Information resulting from a large data discovery service crawling repository landing pages shows that metadata change is common: A study found that 85 \% of dataset landing pages had been changed in the past 5 years \parencite{benjelloun_google_2020}.
DOI metadata are also routinely updated: after 3 years, 80 \% of Crossref metadata records have been changed at least once \parencite{hendricks_crossref_2020}.
Metadata change does not conflict with the registration of DOIs. In its handbook, the DOI foundation emphasizes the relevance of updating DOI metadata: "To maintain quality services, it is also crucial that metadata assigned to a DOI name be regularly updated." \parencite[86]{doi_foundation_doi_2023} Developing a data maintenance policy is the responsibility of DOI registration agencies. The DOI registration agencies DataCite and Crossref are very permissive of changes to metadata records after a DOI is registered.\footnote{DataCite - Can I update the DataCite DOI metadata one the DOI is registered? \url{https://support.datacite.org/docs/should-i-update-the-datacite-doi-metadata-if-the-name-of-the-creator-changes}}\footnote{Crossref - Maintaining your metadata: \url{https://www.crossref.org/documentation/register-maintain-records/maintaining-your-metadata/}}
\\
Although several sources confirm the prevalence of metadata change overall, there is currently little research on metadata change in the context of research data.
Insights from traditional cataloging contexts, such as digital libraries, offer examples of how metadata change can be studied. In these settings, metadata change is considered a normal part of metadata management \parencite{dobreski_descriptive_2021}, and research on the topic has progressed.
\textcite{zavalina_building_2015} developed and tested a framework for metadata change in libraries.
The framework, initially based on Dublin Core metadata, comprises three main categories of metadata change (Addition, Deletion, and Modification). The framework is context-independent and can be adapted to any information object, metadata schema and platform -- in a later study, it was applied to a more complex library cataloging standard \parencite{zavalina_developing_2016}.
\\
The application of the framework to the metadata collection at a university library showed that 42.5 \% of metadata records were changed at least once in the last 5 years \parencite{tarver_how_2014}. Most of these records were changed five times or less, but some outliers were edited more than 50 times. Each year in the observed period, the number of metadata records edited increased. In 93.6 \% of edited metadata records, one or two editors were involved.
Some parts of metadata records tended to be very stable over time, for example values in elements describing language, format, resource type, rights information, publication date, or the title \parencite{zavalina_exploration_2015}. The elements that were changed most frequently were creator, contributor, identifier, and notes.
The authors followed up by comparing the results of their analysis to metadata change in the union catalog WorldCat \parencite{zavalina_metadata_2015}. The two sources differed in the types of changes occurring and the metadata elements edited most frequently.
A longitudinal study of WorldCat metadata records later showed that metadata records were changed extensively over time \parencite{zavalina_quality_2016,zavalina_developing_2016}. After 3 years, all entities in a sample of 369 metadata records compiled in 2013 had been edited at least once, with an average of 4.9 editing events per record.
\\
Metadata change can also be observed outside traditional library contexts. The distribution of change events per metadata records in a collection of digitized patents was uneven, and in most change events, only one metadata element was altered \parencite{zavalina_quality_2017}. Most frequently changed were elements describing the content (subject, description) and entities associated with the object (creator, contributor, publisher) \parencite{zavalina_uncovering_2018}.
\\
In conclusion, there is some research into metadata change in library settings and developments of DOI metadata over time \parencite{van_eck_crossref_2024,hendricks_crossref_2020}, but studies focusing on metadata change in the context of research data are lacking.

\subsection{Metadata versioning and provenance}
The analysis of changes to an object is closely related to the concept of provenance. For research data, provenance involves capturing the complete lineage of a data product, including the context of collection and any transformations performed \parencite{paine_surfacing_2019}.
In most data versioning schemes, metadata change does not create a new version of a dataset \parencite{klump_versioning_2021}. However, as discussed above, metadata records are regularly modified and therefore have a lineage that can be documented. These changes in metadata are often recorded by research infrastructures, but rarely released to the public \parencite{paine_surfacing_2019}. An entity that releases metadata provenance information is the DOI registration service DataCite, which started tracking metadata provenance in March 10, 2019 \parencite{fenner_exposing_2019}.
DataCite metadata provenance information are based on the PROV family of specifications, which were developed by a W3C working group to enable the recording and exchange of provenance information across heterogenous contexts \parencite{moreau_provenance_2013, moreau_prov-dm_2013,moreau_rationale_2015}. Change requests are recorded by DataCite and can be accessed via a dedicated API endpoint.

\section{Methods}
This paper will use DataCite metadata provenance information to analyze the evolution of DataCite metadata records for research data over time. The approach is informed by library and information science research on metadata change that originated in more traditional cataloging settings.
The change types in the framework of metadata change developed by \textcite{zavalina_building_2015} were shown to be flexible enough to describe developments in diverse settings. The same general change types will be used to trace changes in DataCite metadata.
Previous research from research software engineering has demonstrated that provenance information in PROV formats can shed light on change processes of research outputs \parencite{schreiber_analyzing_2021}. Studies of metadata change have shown that provenance information allow for a more detailed analyses of processes, as compared to capturing snapshots of metadata records at fixed intervals \parencite{zavalina_exploration_2015,zavalina_quality_2017,zavalina_uncovering_2018}. Therefore, DataCite metadata provenance information based on PROV will be used in this paper.

\subsection{Data collection}

\subsubsection{Selecting a time frame}
The initial study design intended to look at change events in metadata records created in 2020, allowing for changes to manifest over a time frame of three years. This time frame was chosen because a previous study of Crossref DOI metadata demonstrated that the percentage of metadata records that were changed at least once increased in the first three years after the initial deposit, and flattened considerably after \parencite{hendricks_crossref_2020}.
\\
However, preliminary tests revealed that DataCite metadata provenance information is incomplete up to 2021: first versions of about 20 \% of metadata records could not be accessed via the API. A bug report was filed with DataCite\footnote{DataCite bug report: \url{https://github.com/datacite/datacite/issues/2071} (accessed: 2024-12-03)}.
This issue did not allow a complete or purposeful selection of metadata records created in 2020. Therefore, the study design was adapted to retrieve metadata records created in 2021 and any changes that occurred within 2 years of initial registration.

\subsubsection{Retrieving and parsing data}
Data was collected from DataCite on March 29, 2024.
First, DOIs registered for datasets in 2021 were retrieved from the DataCite API (resourceTypeGeneral = dataset ; created = 2021). In the next step, JSON provenance information was downloaded for these DOIs via the DataCite activities endpoint. Finally, information was extracted from the JSON files. This included the PROV elements, additional administrative information, as well as changes to metadata elements in version 4.4 of the DataCite Metadata Schema. Changes to metadata elements were generally aggregated at the top level in the hierarchy of the DataCite Metadata Schema. These elements are summarized in Table \ref{tab:datacite_4-4}. The metadata element \emph{relatedItem} and child elements were excluded from the analysis, because they are used to describe an object related to the dataset and not the dataset itself.
Version 4.4 was released on March 30, 2021 and is the most current iteration of the DataCite Metadata Schema within the time frame analyzed \parencite{datacite_metadata_working_group_datacite_2021}.
\begin{table}
    \centering
    \begin{tabular}{|p{2.8cm}|p{5.5cm}|p{1.6cm}|p{1.7cm}|p{2.2cm}|}
    \hline
        metadata element & description & occurrence & child elements or attributes & obligation level \\ \hline
        identifier & A unique string that identifies a resource. & 1 & no & mandatory \\ \hline
        creator & The authors involved in producing the resource, in priority order. & 1-n & yes & mandatory \\ \hline
        title & A name or title by which a resource is known. & 1-n & yes & mandatory \\ \hline
        publisher & The name of the entity that holds, archives, publishes prints, distributes, releases, issues, or produces the resource. & 1 & no & mandatory \\ \hline
        publicationYear & The year when the resource was or will be made publicly available. & 1 & no & mandatory \\ \hline
        subject & Subject describing the resource. & 0-n & yes & recommended \\ \hline
        contributor & The institution or person contributing to the development of the resource. & 0-n & yes & recommended \\ \hline
        date & Dates relevant to the resource. & 0-n & yes & recommended \\ \hline
        language & The primary language of the resource. & 0-1 & no & optional \\ \hline
        resourceType & The type of the resource. & 1 & yes & mandatory \\ \hline
        alternateIdentifier & An identifier other than the primary identifier of the resource. & 0-n & yes & optional \\ \hline
        relatedIdentifier & Identifiers of related resources. & 0-n & yes & recommended \\ \hline
        size & Size of the resource. & 0-n & no & optional \\ \hline
        format & Technical format of the resource. & 0-n & no & optional \\ \hline
        version & The version number of the resource. & 0-1 & no & optional \\ \hline
        rights & Rights information for the resource. & 0-n & yes & optional \\ \hline
        description & A description of the resource. & 0-n & yes & recommended \\ \hline
        geoLocation & Spatial information referring to the resource. & 0-n & yes & recommended \\ \hline
        fundingReference & Information about financial support for the resource. & 0-n & yes & optional \\ \hline
        relatedItem & Information about a resource related to the one being registered. & 0-n & yes & optional \\ \hline
    \end{tabular}
    \caption{Metadata elements at the top level in the hierarchy of version 4.4 of the DataCite Metadata Schema; \emph{occurrence} refers to quantity constraints of the elements in the schema}
    \label{tab:datacite_4-4}
\end{table}
The extraction of change information was based on the XML to JSON mapping provided by DataCite \footnote{DataCite XML to JSON mapping: \url{https://support.datacite.org/docs/datacite-xml-to-json-mapping} (accessed: 2024-12-03)}

\subsubsection{Preprocessing}
After extracting the metadata provenance information from DataCite, changes that happened 2 years after a metadata record was first created were removed to ensure that the analysis is based on the same time frame for all metadata records in the sample. Any versions of a metadata record that were registered after that cut-off were excluded from the analysis (211,823 ;  3.42 \%). The final sample comprises 5,985,516 versions of 2,688,310 DataCite metadata records.
\\
In order to analyse how metadata records were changed, changes to metadata elements were grouped into basic categories, similar to the change types in the framework by \textcite{zavalina_building_2015} described above. The categories compare the occurrence of values assigned to a metadata element between two consecutive versions of a metadata record. They cover the general use and types of changes of metadata elements (see Table \ref{tab:categories}). General use includes the categories \emph{unused} and \emph{created}. These categories indicate whether a metadata element has remained unused in the current or previous versions of the metadata record or if a value was entered for the first time in the current version. Change types include \emph{addition}, \emph{deletion}, \emph{modification} and \emph{unchanged}. These categories indicate the ways in which values of a metadata element have been changed (addition or deletion of values; other modifications), or if the values remained unchanged. \emph{Addition} was assigned if new instances of a metadata element were added in the new version of the metadata record, \emph{deletion} if instances were removed, and \emph{modification} if the number of instances remained the same, but the values were changed. Examples of \emph{modification} include instances where spelling errors were corrected or values were updated (for example following name changes).
\begin{table}
    \centering
    \begin{tabular}{|p{2.5cm}|p{3cm}|p{8.5cm}|} \hline
     & category & description \\ \hline
    general use & \emph{unused} & the element has not been used in this or previous versions of the metadata record \\ \hline
    & \emph{created} & the element was first used in this version of the metadata record \\ \hline
    change types & \emph{addition} & a new value was added to the element in this version of the metadata record \\ \hline
    & \emph{deletion} & a value was removed from the element in this version of the metadata record \\ \hline
    & \emph{modification} & the element was otherwise modified in this version of the metadata record \\ \hline
    & \emph{unchanged} & the element was not changed in this version of the metadata record \\ \hline
    \end{tabular}
    \caption{Categories describing the general use of metadata elements and types of changes observed}
    \label{tab:categories}
\end{table}
\noindent
The structure of the DataCite Metadata Schema has a significant impact on the assigned change types. For example, if a metadata element is obligatory, it will most likely be present in the first version of a metadata record, and \emph{deletion} is not possible. If the occurrence of this element is also limited to one instance, \emph{addition} is impossible as well. For the analysis, change types are assigned at the top level in the hierarchy of the DataCite Metadata Schema. As a result, for metadata elements with child elements or attributes, the change type \emph{modification} can also be assigned when child elements or attributes are added or deleted.
\\
For the time series analysis, the time passed between the registration of subsequent versions of metadata records was calculated using the time stamps in the PROV element \emph{generatedAtTime}. A metadata record is connected to the research data repository that registered it via the PROV element \emph{wasAttributedTo}.

\section{Results}

\subsection{Frequency of changes}
\begin{figure}
    \centering
    \includegraphics[width=0.9\linewidth]{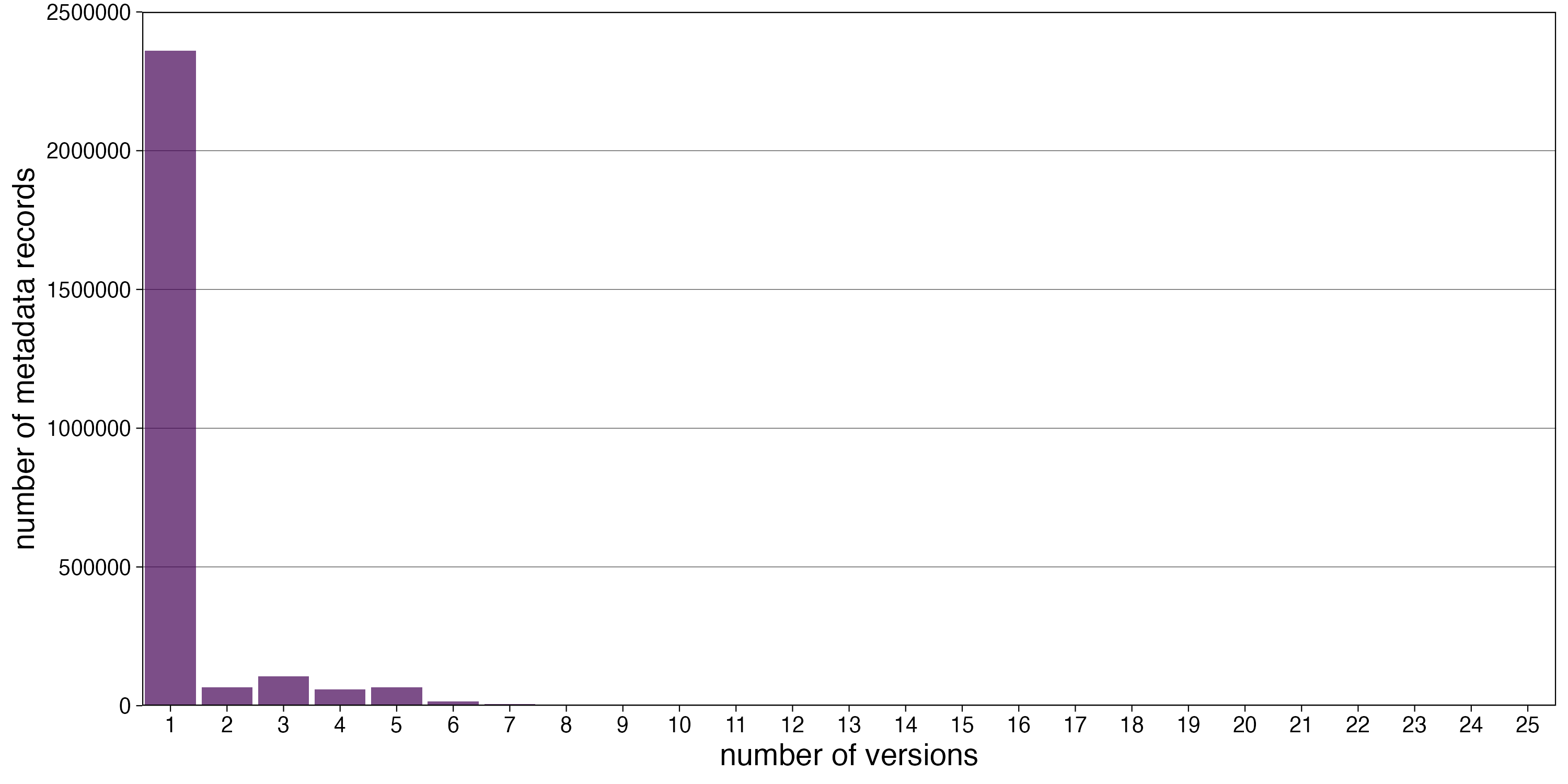}
    \caption{Distribution of the number of changes to a metadata record}
    \label{fig:distribution_changes}
\end{figure}

The results show that 89.05 \% (2,393,969) metadata records in the sample have more than one version.
A closer inspection reveals that in most of these cases, metadata elements in the DataCite Metadata Schema, such as \emph{title} or \emph{publicationYear}, remained unchanged. Instead, additional administrative data captured by DataCite was changed, for example the URL the DOI resolves to.
For 12.18 \% (327564) of metadata records in the sample, elements of the DataCite Metadata Schema were changed at least once within two years of initial registration.
In the following, \emph{metadata change} is defined as changes to metadata records that affect elements in the DataCite Metadata Schema, unless otherwise specified.
\\
Some of all changes (both metadata change and change to administrative information) might be a result of automated processes, for example steps in repositories' metadata management workflows or metadata enrichment. Automated processes are not clearly marked as such in DataCite metadata provenance information. The PROV element \emph{wasAttributedTo} indicates the entity responsible for the recorded changes. In the sample, values in this element are almost exclusively IDs of DataCite clients; only 96 changes are attributed to \emph{admin}, indicating involvement of DataCite. Therefore, PROV \emph{wasAttributedTo} can not be used as an indicator of automated processes. The time passed between edits is also not a clear pointer. Changes that occurred less than five minutes after the previous version of a metadata record was registered by DataCite are similar to other changes in terms of the elements changed and the types of changes enacted (see below).
\\
Only including metadata change, the metadata records in the sample have between 1 and 25 versions (mean = 1.34 ; sd = 1.86) (see Figure \ref{fig:distribution_changes}). The median of 1 indicates that the majority of metadata records were not changed in the observed time period.

\subsection{Metadata elements used in the first and changed in subsequent versions}
\begin{figure}
    \centering
    \includegraphics[width=0.9\linewidth]{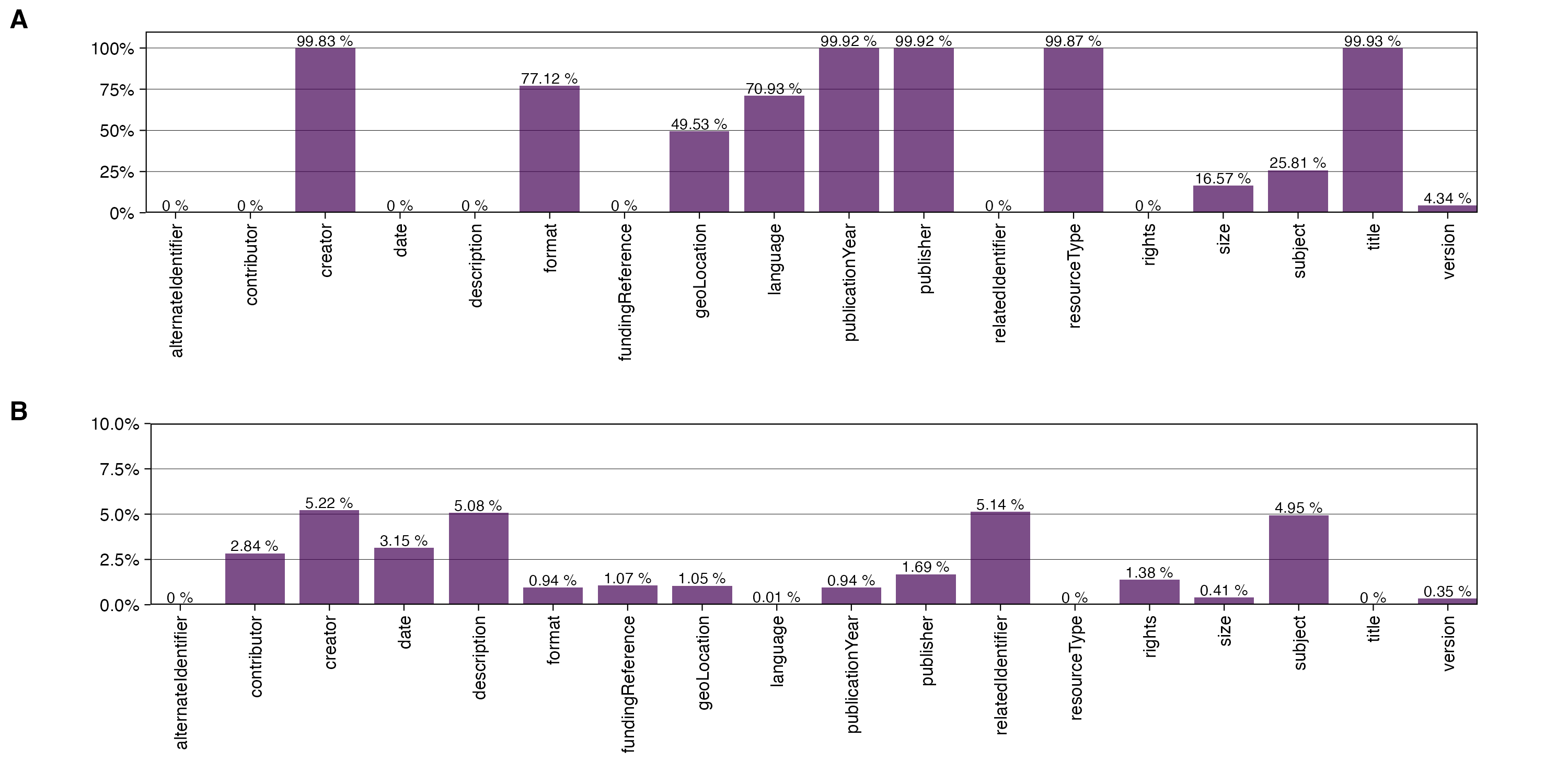}
    \caption{Rate of metadata records that (A) use a metadata element in the first version and (B) change it at least once in later versions}
    \label{fig:use_change_elements}
\end{figure}

Focusing on the first versions of metadata records reveals that overall, the use of elements varies when metadata records are initially registered with DataCite (see Figure \ref{fig:use_change_elements} (A)). Information on the \emph{title}, \emph{publisher}, \emph{publicationYear}, \emph{resourceType}, and \emph{creator} of a dataset are mandatory in the DataCite Metadata Schema and are missing only in rare cases. Present in more than half of newly registered metadata records are statements about the \emph{format} (77.12 \%) and \emph{language} of the dataset (70.93 \%).
\noindent
The analysis shows that metadata elements are changed with varying frequency after they have been initially created (see Figure \ref{fig:use_change_elements} (B)). Most frequently changed are \emph{creator}, \emph{relatedIdentifier}, and \emph{description}, which are changed in 5.22 \%, 5.14 \%, and 5.08 \% of metadata records in the sample, respectively. Changes are rare for \emph{title} (0 \%), \emph{language} (0.01 \%), \emph{version} (0.35 \%), \emph{size} (0.41 \%), \emph{publicationYear} (0.94 \%), and \emph{format} (0.94 \%) are rare. Changes to the elements \emph{resourceType}, and \emph{alternateIdentifier} are not present in the sample.
On average, statements in 1.99 elements are changed in revisions of the metadata records in the sample.

\subsection{Types of changes}
\begin{figure}
    \centering
    \includegraphics[width=0.9\linewidth]{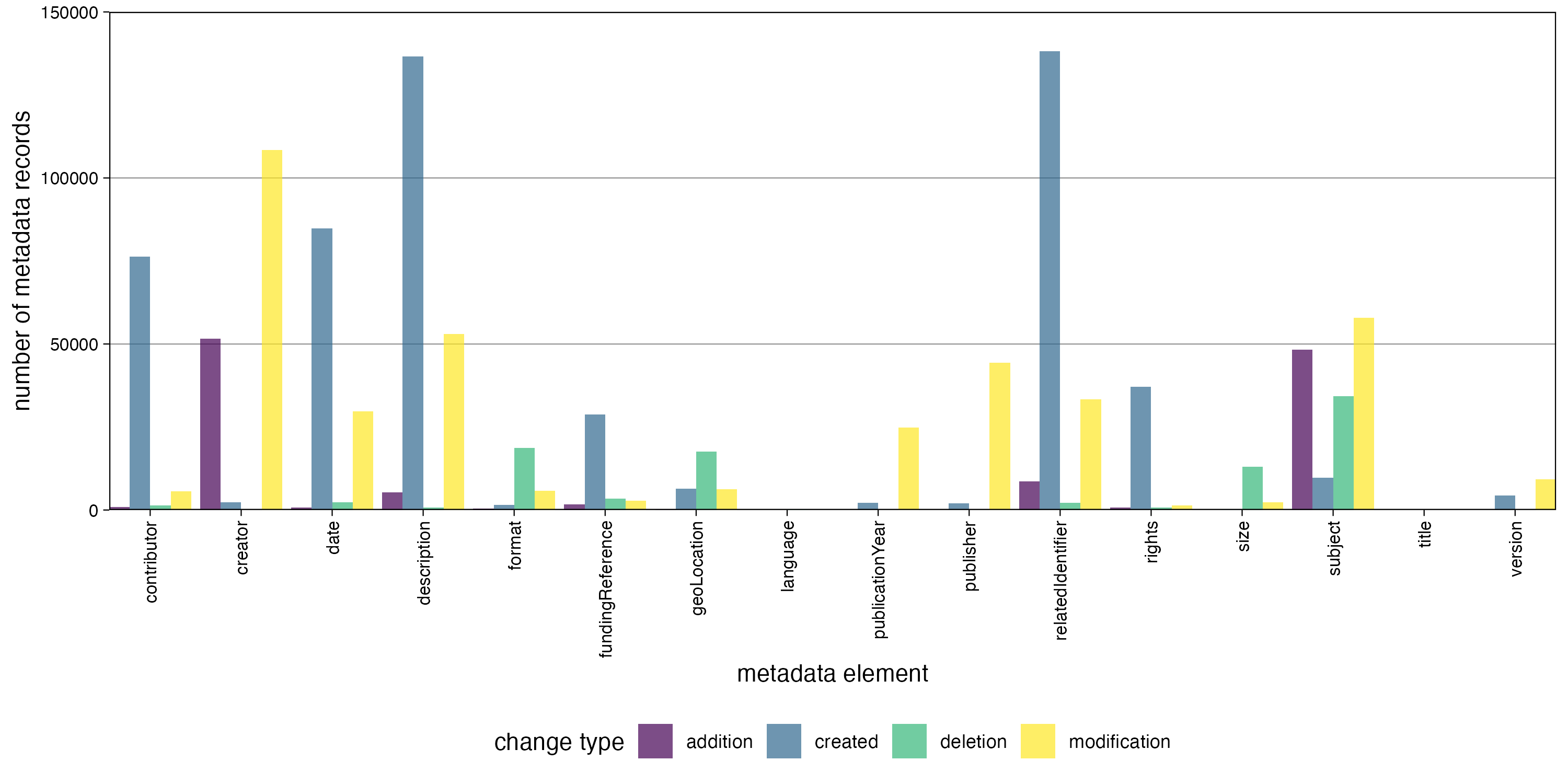}
    \caption{Types of changes by metadata element}
    \label{fig:change_types_elements}
\end{figure}

The most common type of change when considering all metadata elements is \emph{created} (530,371), followed by \emph{modification (385,160),} \emph{addition} (118,298) and \emph{deletion} (94,512).
The frequency of change types observed differs across the individual metadata elements (see Figure \ref{fig:change_types_elements}). Created is most common for the metadata elements \emph{relatedIdentifier}, \emph{description} and \emph{date}, and least common for \emph{language}, \emph{title} and \emph{size}.  Modification is most common for \emph{creator}, \emph{subject} and \emph{description}, and least common for \emph{size}, \emph{rights} and \emph{title}. Addition is most common for \emph{creator}, \emph{subject} and \emph{relatedIdentifier}, and least common for \emph{publicationYear}, \emph{title}, and \emph{version}. Deletion is most common for \emph{subject}, \emph{format} and \emph{geoLocation}, and least common for \emph{language}, \emph{version} and \emph{creator}.
Overall, 11.7 \% (315,170) metadata records were expanded in some capacity after they were first created. This includes instances where at least one metadata element was added or newly created after the first version of the metadata record was initially registered (categories \emph{created} and \emph{addition} in table \ref{tab:categories}).
However, the increase in the number of metadata elements used between the first and last version is low. On average, the metadata records in the sample that were changed only added 0.22 previously unused metadata elements in the observed time period.

\subsection{Time series analysis}
\begin{figure}
    \centering
    \includegraphics[width=0.9\linewidth]{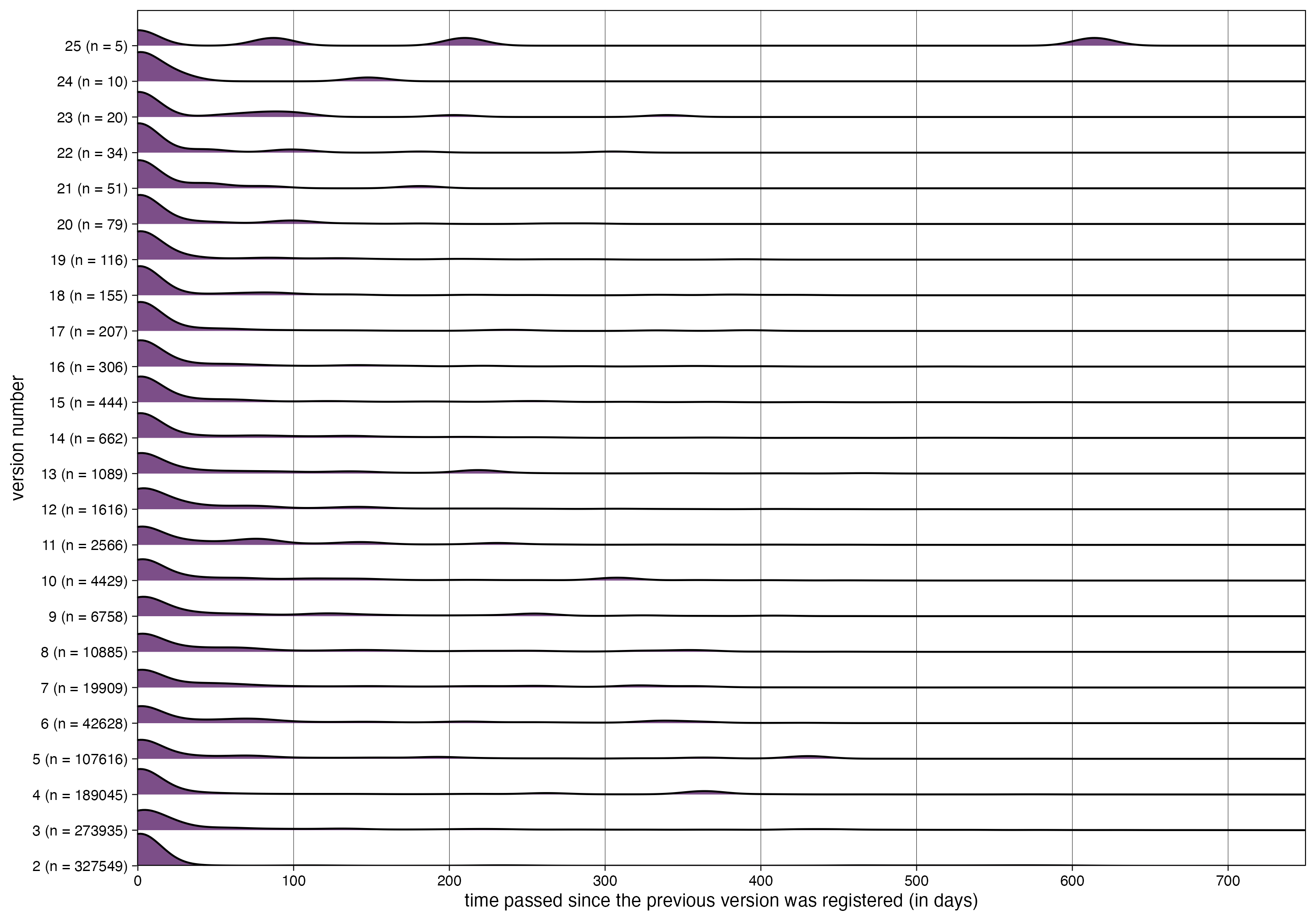}
    \caption{Distribution of the time passed between versions of metadata records in days}
    \label{fig:time_series}
\end{figure}

\begin{figure}
    \centering
    \includegraphics[width=0.9\linewidth]{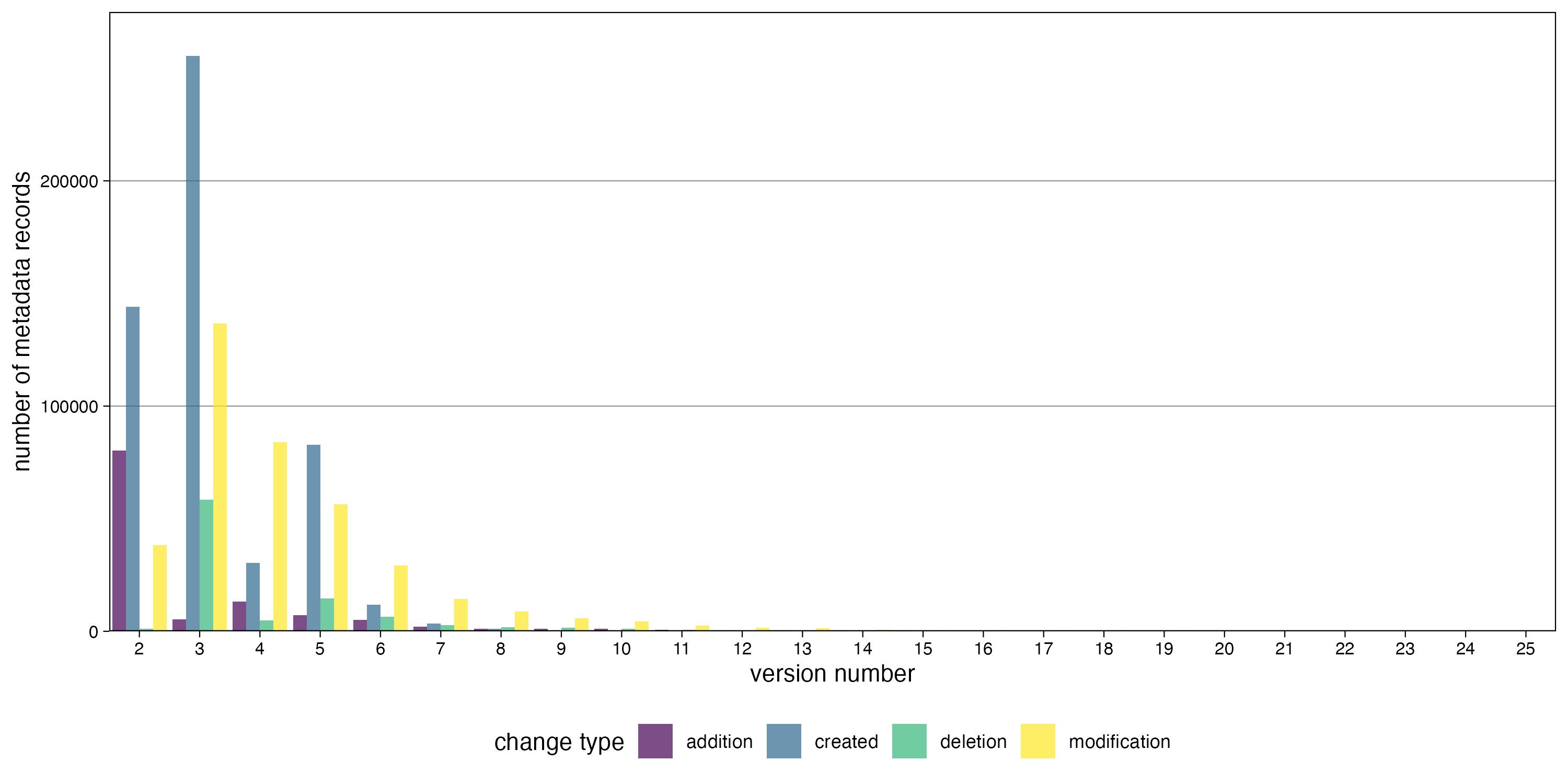}
    \caption{Types of changes by version of metadata record}
    \label{fig:change_types_version}
\end{figure}

The analysis of the time passed between the registration of subsequent versions of metadata records reveals that on average, a new version is released after 77 days (median: 1.9 days). Figure \ref{fig:time_series} shows the distribution of the time passed between versions of metadata records. It shows that the distributions skew left for most versions, indicating that changed metadata records tend to be released after a relatively short time.
\\
At DataCite, there are three DOI states\footnote{DataCite DOI states: \url{https://support.datacite.org/docs/doi-states} (accessed: 2024-12-03)}, and only once the state of a metadata record is set to \emph{findable} is it indexed in DataCite and can be accessed publicly. Once the state is set to \emph{findable}, it can no longer be reverted to the \emph{draft} state.
77.18 \% (2,074,957) of the metadata records in the sample are created in the \emph{findable} state. The other metadata records are changed to be findable in periods ranging from less than a minute to over 3 years, with an average of 7.6 days and a median of 0.01 minutes.
\\
In the first change of a metadata record (version 2), \emph{created} and \emph{addition} are the most common change types observed. Overall, \emph{created} is most common in version 3. After that, the frequency of both types of extending metadata records (\emph{created} and \emph{addition}) drop off and modification becomes the most common type of change (see Figure \ref{fig:change_types_version}). Deletions most frequently occur when a metadata record is changed for the second time (version 3).
\\
Metadata elements are first added to a metadata records (category \emph{created} in table \ref{tab:categories}) at different points in time (see Table \ref{tab:time_series_elements}). The metadata elements \emph{creator}, \emph{format}, \emph{geoLocation}, \emph{language}, \emph{publicationYear}, \emph{publisher}, \emph{resourceType}, \emph{size}, \emph{subject}, \emph{title} and \emph{version} tend to be included in the first version of the metadata record. Some of these elements are mandatory and must be submitted to DataCite upon DOI registration. The elements \emph{contributor}, \emph{date}, \emph{description}, \emph{fundingReference}, \emph{relatedIdentifier}, \emph{rights} tend to be added to the metadata records in later versions.
\begin{table}
    \centering
    \begin{tabular}{|l|r|r|r|r|r|}
    \hline
        ~ & ~ & version number & ~ & time passed (in days) & ~ \\ \hline
        metadata element & n & median & mean & median & mean \\ \hline
        contributor & 76270 & 3 & 2.76 & 326.65 & 287.18 \\ \hline
        creator & 2686079 & 1 & 1 & 0 & 0.03 \\ \hline
        date & 84781 & 5 & 3.91 & 52.76 & 146.82 \\ \hline
        description & 136572 & 3 & 3.1 & 44 & 130.68 \\ \hline
        format & 2074797 & 1 & 1 & 0 & 0.05 \\ \hline
        fundingReference & 28734 & 2 & 2.79 & 474.09 & 386.76 \\ \hline
        geoLocation & 1337748 & 1 & 1.01 & 0 & 0.31 \\ \hline
        language & 1906742 & 1 & 1 & 0 & 0 \\ \hline
        publicationYear & 2688298 & 1 & 1 & 0 & 0.02 \\ \hline
        publisher & 2688303 & 1 & 1 & 0 & 0.02 \\ \hline
        relatedIdentifier & 138170 & 3 & 3.37 & 61.5 & 156.06 \\ \hline
        resourceType & 2684862 & 1 & 1 & 0 & 0 \\ \hline
        rights & 37039 & 3 & 3.21 & 264.98 & 258.38 \\ \hline
        size & 445565 & 1 & 1 & 0 & 0.09 \\ \hline
        subject & 703542 & 1 & 1.02 & 0 & 1.4 \\ \hline
        title & 2686347 & 1 & 1 & 0 & 0 \\ \hline
        version & 121094 & 1 & 1.09 & 0 & 3.66 \\ \hline
    \end{tabular}
    \caption{Chronological development of metadata records: Median and mean version number when an element is first created as part of a metadata record; median and mean time passed between the initial creation of the metadata record and the creation of the metadata element (in days)}
    \label{tab:time_series_elements}
\end{table}

\subsection{Patterns at research data repositories}
\begin{figure}
    \centering
    \includegraphics[width=0.9\linewidth]{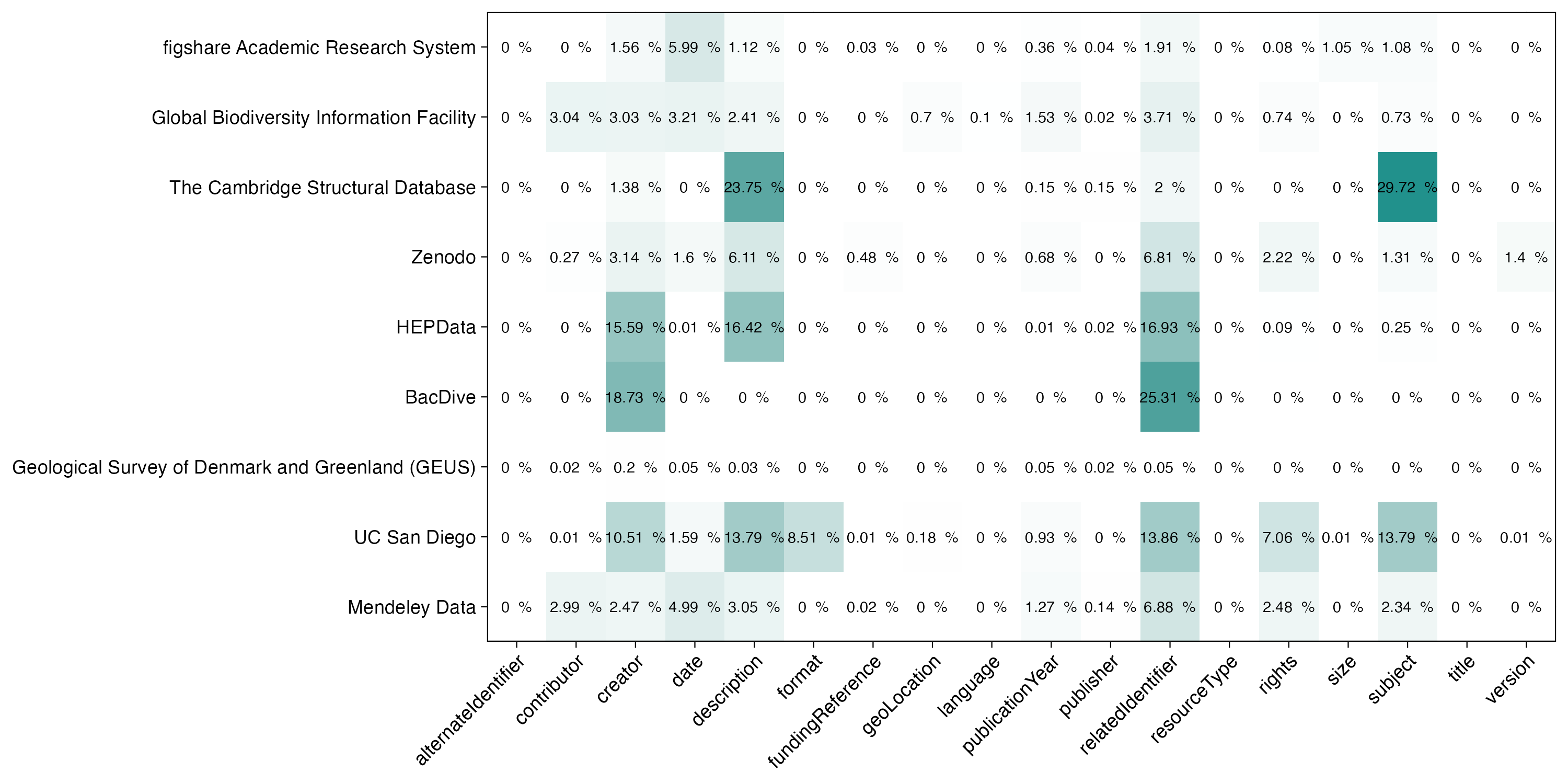}
    \caption{Percentage of metadata records in the collection of a research data repository in which a specific metadata element was changed}
    \label{fig:repositories_elements}
\end{figure}

Patterns of change can be broken down to individual research data repositories using the PROV element \emph{wasAttributedTo}. Changes are recorded at the level of DataCite clients, not at the level of individuals changing metadata records. 777 research data repositories are represented in the sample.
Between 0 and 26.83 \% of metadata records in the collections of these repositories have been changed at least once.
The 10 research data repositories that are responsible for registering the most metadata records in the sample show varying tendencies to adjust specific metadata elements. At the largest repository in the sample, Plutof, no metadata change was observed (only changes to administrative information). In contrast, at 5 other repositories in this group, 10 or more metadata elements have been changed at least once in the entire collection of metadata records. Figure \ref{fig:repositories_elements} shows the percentage of metadata records in the entire collection of a the largest research data repositories (excluding Plutof) in which a specific metadata element was changed. It reveals that some metadata elements are changed regularly at individual research data repositories. For example, \emph{subject} and \emph{description} are changed frequently at the Cambridge Structural Database \footnote{Cambridge Structural Database: \url{https://doi.org/10.17616/R36011}}, and \emph{creator} and \emph{relatedIdentifier} at BacDive\footnote{BacDive: \url{https://doi.org/10.17616/R31NJMKK}}.
\emph{Created} is the most frequently observed type of metadata change at Mendeley, GBIF, Zenodo, HEPData, and figshare Academic Research System, \emph{modification} at BacDive, Geological Survey of Denmark and Greenland (GEUS), and UC San Diego, and \emph{addition} at The Cambridge Structural Database.

\section{Discussion}

\subsection{Fixity and fluidity of metadata records}
The analysis reveals that most DataCite metadata records describing research data have more than one version. However, in most of these cases, only administrative data, such as the URL the DOI resolves to, was changed. Metadata elements in the DataCite Metadata Schema were changed for  12.18 \% (327564) of metadata records within two years of initial registration.
\\
Most metadata records remain unchanged, and most changes are small -- on average, only 1.99 metadata elements are changed compared to the previous version.
Most changes introduce metadata elements that were not present in previous versions, followed by the modification of existing values. The addition and deletion of metadata elements at the top level in the hierarchy of the DataCite Metadata Schema is comparatively rare. This indicates that metadata change is incremental; changes to DataCite metadata records tend to be targeted on few specific elements, rather than a complete overhaul of the entire metadata record.
Some of the changes observed might be the result of automated processes, but automated and manual interventions can not be distinguished reliably.
\\
An interesting result of the analysis is that even elements that could be considered relatively fixed are changed, for example \emph{creator} or \emph{subject}. Future research could look at these changes in more detail, for example to determine if changes are a result of correcting errors, or if values are changed more profoundly. Stakeholders in the DOI registration process could also discuss if profound changes in metadata warrant a new dataset version.
\\
Overall, the analysis shows that DataCite metadata records balance fixity and fluidity. In general, DataCite metadata records can be changed, but are stable over time, resulting in reliable descriptions of research data. Metadata elements describing technical details of datasets, such as \emph{size} or \emph{format}, are among the most stable metadata elements. At the same time, there is some degree of mutability, allowing for adaptions as needed.
\\
DataCite is the largest source of information about research data, and it is a provider of open research information in the sense of the Barcelona Declaration on Open Research Information.
The relative stability of DataCite metadata records means that they can be reasonably used in scientometric analyses. This means that in the context of research data, a reliance on proprietary bibliographic databases could be avoided.
\\
For datasets that are expected to change frequently, it might be advisable to state the version of the metadata records used in scientometric studies.
\\
Changes in a metadata record might be the result of the publication of a new version of the dataset it describes. For example, the modification of the metadata element \emph{version} can be an indicator of dataset versioning, as well as additions or modifications of \emph{relatedIdentifier}. However, more research is needed to determine the degree to which dataset versioning and metadata versioning are coupled.

\subsection{Comparison with metadata change in other contexts}
Compared to metadata change in traditional cataloging settings, metadata change at DataCite is less common, but they share some characteristics. In both environments, the number of changes is distributed unequally across metadata records. Some metadata elements also tend to be more stable over time than others, both in traditional cataloging settings and at DataCite. For example, information on creators of a resource are changed relatively frequently both at DataCite and in traditional cataloging environments \parencite{zavalina_exploration_2015}.
\\
Compared to observations made at another DOI registration agency, Crossref \parencite{hendricks_crossref_2020}, the rate of changes to DataCite metadata records is lower. Identifying the cause of the difference is outside the scope of this analysis, but it could be a result of not distinguishing between metadata change and changes of administrative information.
\\
Previous research observed that of 24.7 million dataset landing pages, 85 \% had been updated within 5 years \parencite{benjelloun_google_2020}. The rate of changed metadata records at DataCite is much lower, which could indicate that changes made to metadata records at research data repositories locally do not translate well to DataCite. This is also an issue in the wider scholarly communications landscape. The Collaborative Metadata Enrichment Taskforce (COMET) investigates options for improving metadata flows between metadata providers and metadata users \parencite{buttrick_comet_2025}. This study indicates that DataCite metadata records are updated infrequently. Therefore, there is potential for improving metadata quality by feeding metadata enriched by harvesters back to DataCite.
\\
Going forward, more research is needed to better understand how metadata records are created and managed at research data repositories, what information is passed on to DataCite, and what information gets lost along the way. Systematic comparisons between metadata practices in different environments could also be useful to identify characteristics of metadata practices that are specific to research data.

\subsection{Metadata maintenance as a continuous task}
In theory, metadata maintenance should be a continuous task -- changes in the environment or user base of a repository should lead to changes in metadata \parencite{aspray_talking_2019}. To some extent, the analysis supports this, since change to DataCite metadata records after publication does occur. However, the frequency of changes metadata records and the time passed between changes demonstrates that continuous and regular change is not a common practice overall. This observation could change when analysing changes to metadata over a longer time period -- two years might not be enough time for impactful changes to manifest in the environment of a repository and later in DataCite metadata records.

\subsection{Metadata change and metadata completeness}
Due to the limited time period observed, the sample likely does not cover the full extent of change to metadata records. Despite this restriction, there is some indication that the observed changes contribute to metadata completeness: 11.7 \% of metadata records in the sample were made more comprehensive by changes implemented after the initial registration. However, these changes rarely expand descriptions of research data: On average, only 0.22 previously unused metadata elements were added to metadata records in the observed time period.
\\
With the chosen method of aggregating changes to metadata elements at the top level in the hierarchy of the DataCite Metadata Schema, interventions that result in increased completeness of metadata records can be found not only in change types \emph{created} (a top-level metadata element that was not present in previous versions is now used) and \emph{addition} (a new instance of a top-level metadata element that was present in previous versions is added), but also in some cases in \emph{modification} -- if new sub-elements and attributes are added to a top-level element already in use. Future research could investigate metadata change on a more granular level to determine the full effect metadata change has on the completeness of DataCite metadata records and identify targeted strategies to incentivise DataCite clients to expand descriptions of research data.

\subsection{Metadata practices at research data repositories}
A first look at the largest repositories in the sample shows some indication of distinct metadata practices at individual research data repositories: Some repositories tend to change certain metadata elements more than others. This might be related to general characteristics of a repository, such as its type or scope, or it might be a result of specific practices manifested in workflows, guidelines or automated processes.
\\
Future research could look further into metadata practices at individual research data repositories and how metadata travel from these distributed sites to central sites of aggregation, such as DataCite. A more detailed understanding of these factors could inform targeted interventions by DataCite to improve metadata quality or foster cooperation between repositories with similar collections or challenges.

\section{Conclusion}
This paper examined the overall prevalence and patterns of change to DataCite metadata records. It is the first to use metadata provenance information to study changes in DataCite metadata records.
\\
Metadata change was observed for 12.18 \% of metadata records in the sample, and change tends to be incremental and not extensive. DataCite metadata records offer reliable descriptions of datasets and are stable enough to be used in scientometric research.
\\
The rate of change differs from previous studies of metadata change in other contexts, suggesting that there are differences in metadata practices between research data repositories and more traditional cataloging environments.
\\
The observed changes do not seem to fully align with idealized conceptualizations of metadata creation and maintenance for research data. In particular, the data does not show that metadata records are maintained routinely and continuously. Metadata change also has a limited effect on metadata completeness.
\\
The relative stability of DataCite metadata records means that they can be reasonably used in scientometric analyses. Because DataCite is the largest source of information about research data, and a provider of open research information, a reliance on proprietary bibliographic databases could be avoided in the context of research data.
\\
The study found first evidence of distinct metadata practices at individual research data repositories. Future research could investigate metadata practices at research data repositories to get a more realistic view of processes and derive feasible approaches to improve metadata quality.

\section*{Limitations}
Due to technical difficulties, this study is limited to changes to DataCite metadata records that occurred within 2 years of the DOI being first registered. Therefore, the results do not permit any inferences about changes that may occur over a longer period of time.

\section*{Data availability}
The data is published in a repository under an open license.
\\
Strecker, Dorothea (2024) Changes in DataCite DOI metadata for research data. Version 1.0: \href{http://doi.org/10.5281/zenodo.14274240}{DOI: 10.5281/zenodo.14274240}

\printbibliography

\theendnotes

\end{document}